\documentclass{eas}
\usepackage{graphicx}
\begin{document}

\title{From the atmosphere to the circumstellar environment in cool evolved stars} 
\runningtitle{Cool evolved stars}
\author{M. Wittkowski}\address{ESO, Karl-Schwarzschild-Str. 2, 85748 Garching bei M\"unchen, Germany}
\author{C. Paladini}\address{Institut d'Astronomie et d'Astrophysique,  Universit{\'e} Libre de Bruxelles, ULB, CP.226,  Boulevard du Triomphe,  1050 Brussels, Belgium}
\begin{abstract}
We discuss and illustrate contributions that optical interferometry 
has made on our current 
understanding of cool evolved stars. We include red giant branch (RGB) stars, 
asymptotic giant branch (AGB) stars, and red supergiants (RSGs).
Studies using optical interferometry from 
visual to mid-infrared wavelengths have greatly increased our knowledge of
their atmospheres, extended molecular shells, dust formation, and winds.
These processes and the morphology
of the circumstellar environment are important for the further evolution
of these stars toward planetary nebulae (PNe) and core-collapse supernovae
(SNe), and for the return of material to the interstellar medium.
\end{abstract}
\maketitle
\section{Introduction}
A stellar atmospheric model describes the temperature and
density stratification of the atmosphere, i.e. the variation
of temperature and pressure with height in the atmosphere.
In addition to the vertical stratification there may be 
horizontal temperature and density inhomogeneities
due to, e.g., rotation, magnetic fields, or convection.
Together with opacities, these stellar model atmospheres
predict the spectrum emerging from every point of a stellar 
disk. Most often model atmospheres are calibrated by and
compared to integrated stellar spectra.
However,
spatially resolved observations can measure the intensity
profile across the stellar disk and can thus provide stronger
constraints on model atmospheres. Such observations have been
successfully executed by direct imaging observations of the sun, 
and by interferometric observations for the case of stars other 
than the sun.

Cool evolved stars finish their evolution that is driven 
by nuclear fusion along the Hayashi track in the Hertzsprung Russel (HR)
diagram. They are characterized by low effective temperatures
between about 2500\,K and 4000\,K and low surface 
gravities.  They show extended atmospheres that have conditions
favorable for the formation of molecules and dust. 
Dust is accelerated dragging
along the gas, a process that leads to circumstellar shells
and the expulsion of gas and dust into the interstellar medium via
winds. From here on, the further stellar evolution of asymptotic
giant branch (AGB) stars (initial masses up to about 8 solar masses)
toward planetary nebulae (PNe) and of red supergiants (RSGs, initial masses above
about 8 solar masses) toward
core-collapse supernovae (SNe) is mostly driven by the mass-loss process.
The details of this process are currently best constrained for carbon-rich
AGB stars, i.e. stars for which carbon has been dredged up from the core 
into the atmosphere, as outlined by, e.g., Wachter \etal\ (\cite{wachter2002})
and Mattsson \etal\ (\cite{mattsson2010}). The process is less constrained
for oxygen-rich AGB stars (e.g., Woitke \cite{woitke2006}, 
H\"ofner \cite{hoefner2008}) and for red supergiants 
(e.g., Josselin \& Plez \cite{josselin07}).
In order to constrain and to understand these processes, it is important to
observationally establish the detailed stratification and geometry of the
extended atmosphere and the dust formation region, and to compare it to
different modeling attempts. Here, interferometry plays a crucial role
by its ability to spatially resolve the atmospheres and innermost regions
of the dust formation zone, where the mass-loss process is initiated.

A few selected examples of interferometric observations of stellar 
atmospheres and circumstellar environments
include constraints of the limb-darkening effect (e.g., Quirrenbach
\etal\ \cite{quirrenbach96}, Wittkowski \etal\ \cite{wittkowski2001,wittkowski2004}),
comparisons with models
of photospheric convection (e.g., Chiavassa \etal\ \cite{chiavassa2010}),
interferometric observations of oblate shapes of rotating stars
(e.g., Domiciano de Souza \cite{deSouza2003}), constraints of stellar
pulsations of Cepheids (e.g., Kervella \etal\ \cite{kervella2004})
and Mira variables (e.g., Thompson \etal\ \cite{thompson2002}), 
dust formation around Cepheids (e.g., Kervella \etal\ \cite{kervella2006})
and Miras (e.g., Wittkowski \etal\ \cite{wittkowski2007}), observations
of disks in binary post-AGB stars (e.g., Deroo \etal\ \cite{deroo2007}),
or investigations of the central regions of PNe ( e.g., Chesneau \etal\
\cite{chesneau2007}).

In the following section we describe in more detail interferometric
observations of stellar atmospheres on the (first) red giant branch (RGB), 
the AGB, and of RSGs.
We continue with dust shells and winds from these cool evolved stars,
and finally describe recent observational results on asymmetries 
of their circumstellar dust environments.
In the context of the VLTI school book, we focus on results obtained
with the VLTI, but also include examples of important results
from other interferometers.
Evolutionary stages preceding and succeeding RGB, AGB, and RSG
stars are not (fully) covered by this article, as for instance hotter
giants and supergiants, post-AGB stars, PNe, or Wolf-Rayet stars.

\section{Stellar atmospheres of cool evolved stars}
Interferometric observations of stars provide to first order
stellar radii and such measurements have been the prime scientific
target of early interferometers. The next-order effect of
resolving a stellar surface is a measurement of the limb-darkening
effect, where stars appear brighter at the center of the stellar
disk and darker at the limb of the stellar disk.
This phenomenon is a consequence of the vertical temperature
stratification of the stellar atmosphere combined with the
line-of-sight of the observation. Along any line of sight
the observer will see into a geometrical depth of the atmosphere 
that corresponds to an optical depth of unity. 
Toward the center of the star, deeper atmospheric layers are seen, which are
hotter because of the vertical temperature stratification; toward
the limb of the star shallower atmospheric layers are seen, which are cooler.
Observations of this effect thus constrain the vertical temperature
stratification of the atmosphere, which is one of the prime outcomes
of a model atmosphere.
The strength of the effect depends on the temperature stratification
of the star and on the wavelength of observation.

Model atmospheres are based on different model geometries, plane-parallel
geometries and spherical geometries. The plane-parallel model is semi-infinite
for all viewing angles, i.e. all paths are optically thick. It has a
singularity at a viewing angle of 90 deg., where the intensity drops suddenly
to zero. The spherical model has an optically thin limb. This causes
an inflection point of the intensity profile and a tail-like extension.
More details on the effects of model geometries can be found in Sect. 3.4
of Wittkowski \etal\ (\cite{wittkowski2004}).
Observationally, effects of the true spherical nature of stars are
more pronounced for stars that have very extended atmospheres, as in the
case of cool evolved stars discussed here.

It has been observationally established that the photospheres of 
AGB stars and red supergiants are surrounded by atmospheric molecular layers
(e.g., Mennesson \etal\ \cite{mennesson2002}, 
Perrin \etal\ \cite{perrin2004,perrin2005}). In these
cases the model geometry is more complex than a simple plane-parallel
or spherical geometry. Instead, the intensity profiles are strongly
wavelength dependent and show multiple components or strong 
tail-like extensions (cf. Scholz \cite{scholz2003}).

\subsection{Radius definitions and fundamental stellar parameters}
The definition of a stellar radius is not easy because of the complex
geometries of the stellar atmospheres, as described above, and the complex
resulting intensity profiles, cf. Scholz (\cite{scholz1997}).
This is in particular true for cool evolved stars, which show extended
atmospheres and extended molecular layers.

Different definitions have stellar diameters have been used 
in the literature in the context of interferometric observations:
\begin{itemize}
\item{\it Uniform disk radius:} A uniform disk (UD) diameter is obtained
from a fit of a uniform disk model to the measured visibility data, i.e.
determining to first order the radius of a uniform disk that has the
same integral flux as the true intensity profile. Since the true
intensity profile is not a uniform disk but limb-darkened or
even more complex (see above) and depends on the wavelength, 
this uniform disk will be wavelength-dependent and does not correspond
to a certain physical layer of the star. However, it can be corrected
to a limb-darkened or Rosseland-mean (see below) radius using model
atmospheres.
\item{\it 0\% intensity radius:} The 0\% intensity radius corresponds to the
radius where the intensity drops to zero. This is the radius that is
obtained by fitting a model atmosphere to interferometric data.
However, for a spherical geometry, the intensity never drops to zero
but is set to zero at an arbitrary outer boundary condition.
It can be corrected to a Rosseland-mean (see below) radius using 
model atmospheres.
\item{\it 50\% radius or FWHM:} A 50\% radius or a full width half 
maximum (FWHM) diameter
corresponds to the radius at which the intensity drops to 50\%
of its maximum value. This radius is better defined than a UD radius
or a 0\% intensity radius in the case of very extended intensity profiles.
It depends on the bandpass in the same way as the UD diameter.
\item{\it Rosseland radius:} The Rosseland radius corresponds to the 
atmospheric layer where the Rosseland-mean optical depth equals 2/3.
This definition is physically most meaningful, as it is independent of
wavelength and corresponds to the radius as used in defining
physical quantities such as the luminosity or the effective temperature.
However, it is not a direct observable but has to be derived from observations
based on model atmospheres. It may also be contaminated by extended 
molecular and dusty layers that are located above the photosphere but
contribute to the Rosseland-mean opacity. It may observationally best
be determined by comparisons of observations in near-continuum bandpasses 
together with model atmosphere predictions.
\end{itemize}
A physically meaningful determination of a stellar radius 
is also important
to derive fundamental stellar parameters, most importantly  
effective temperature and luminosity.
Several teams used the
VLTI to concentrate on accurate measurements of stellar radii and 
fundamental parameters of cool evolved stars, including
Richichi \& Roccatagliata (\cite{richichi2005}), 
Wittkowski \etal\ (\cite{wittkowski2006b}), 
Cusano \etal\ (\cite{cusano2012}),
Cruzal\'ebes \etal\ (\cite{cruzalebes2013}), 
Arroyo-Torres \etal\ (\cite{arroyo2013,arroyo2014}),
Paumard \etal\ (\cite{paumard2014}).

\subsection{Red giants and the limb-darkening effect}
\label{sec:RGB}

\begin{figure}
\centering
\includegraphics[width=0.495\hsize]{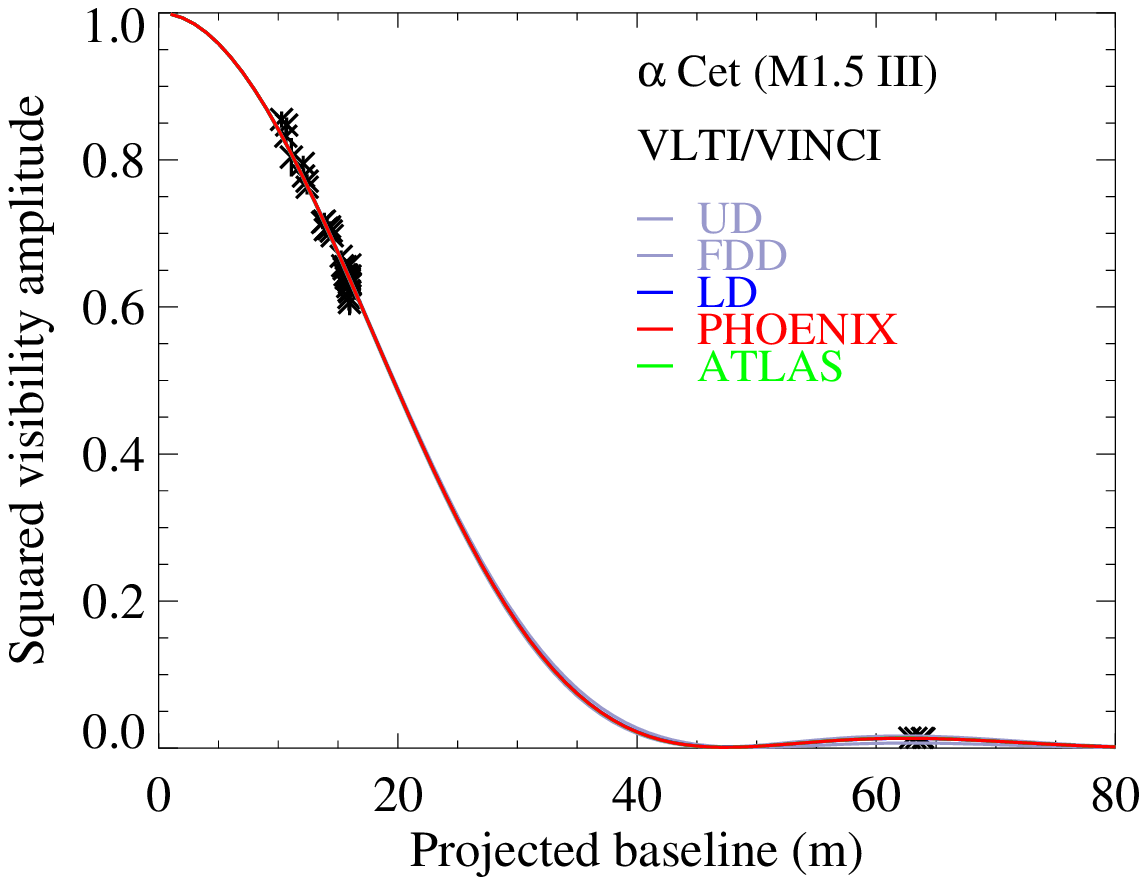}
\includegraphics[width=0.495\hsize]{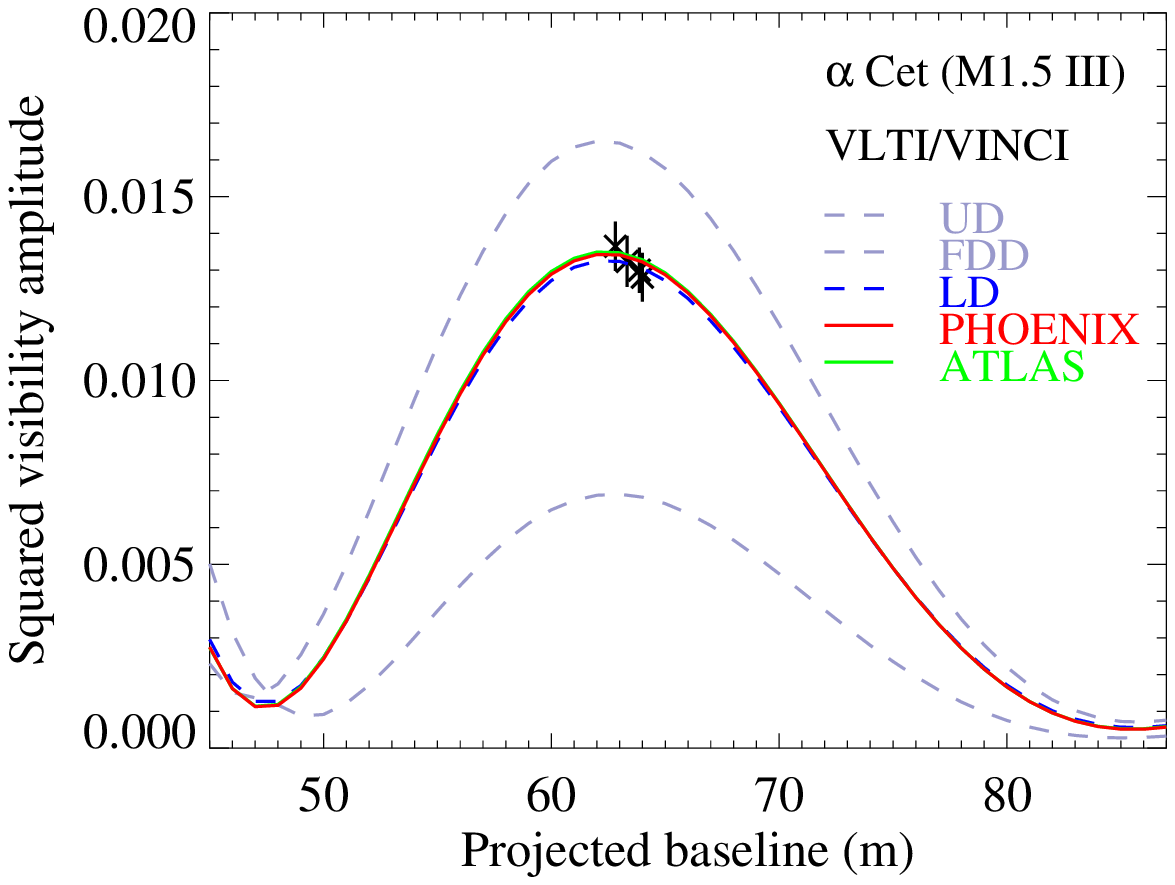}
\caption{VLTI/VINCI observation of the limb-darkening effect
of the red giant Menkar ($\alpha$~Ceti). The left panel shows
all obtained visibilities, and the right panel an enlargement
of the low visibilities within the 2nd visibility lobe. From 
Wittkowski \etal\ (\cite{wittkowski2006a}).}
\label{fig:menkar}
\end{figure}

Measurements of stellar radii using for instance a uniform disk approximation
can be obtained even if the disk is not fully resolved, i.e. by visibility 
measurements within the first lobe.
Direct observations of the limb-darkening effect require a higher spatial
resolution that fully resolves the stellar disk, i.e. requires visibility
measurements within the second lobe.  These are more difficult to obtain
because they require longer baselines and -more importantly- they require
observations of low visibilities corresponding to vanishing fringe
contrasts. 
Early observations of the limb-darkening effect of red giants were
obtained by Quirrenbach\etal\ (\cite{quirrenbach96}) with the
MARK III interferometer, Hajian \etal\ (\cite{hajian1998}) and 
Wittkowski \etal\ (\cite{wittkowski2001}) with the NPOI interferometer,
and by Wittkowski \etal\ (\cite{wittkowski2004, wittkowski2006a}) with the VLTI.

There are in principle two ways to probe the limb-darkening effect:
\begin{itemize}
\item Interferometric observations in the 2nd lobe of the visibility
function at one bandpass. This method directly probes the
shape of the intensity profile.
\item Measuring the variation of a (UD) diameter versus wavelength.
This method
probes the wavelength-dependent strength of the limb-darkening effect.
\end{itemize}
Model atmospheres are best constrained using a combination of both
methods, i.e. by observations in the 2nd lobe of the visibility
function at several bandpasses.

Fig.~\ref{fig:menkar} shows as an example a measurement of the 
limb-darkening effect of the M giant Menkar ($\alpha$~Ceti) obtained
with the VINCI instrument of the VLTI by 
Wittkowski \etal\ (\cite{wittkowski2006a}). It shows that the uniform
disk model and the model atmosphere prediction are virtually identical 
within the first lobe of the visibility function and differ only in the 
second lobe. The maximum of the visibility within the second lobe
is then a measure of the strength of the limb-darkening effect.
Comparisons of different models and different bandpasses show
a trend that the limb-darkening effect is stronger at shorter
(visual) wavelengths and weaker (i.e. closer to the UD prediction)
at longer (infrared) wavelengths and that it is stronger for 
cooler stars, which have more extended atmospheres, than for hotter stars.

Finally, it should be noted that 3-dimensional (3-d) model atmospheres including effects 
of photospheric convection predict a slightly different strength of the 
limb-darkening effect compared to 1-dimensional (1-d) models of the same stellar 
parameters. Hotter, rising granules have a warmer temperature structure
than cooler, descending dark lanes. The mean temperature structure
of a 3-d model then differs from that of a 1-d model
(e.g. Hayek \etal\ \cite{hayek2012}). This effect has been
observed interferometrically by Aufdenberg \etal\ (\cite{aufdenberg2005})
for Procyon. High-precision interferometric observations in continuum
bandpasses may thus in future
provide further constraints on stellar convection and 3-d model atmospheres.

\subsection{Atmospheres of AGB stars}
AGB stars are affected by stellar pulsations starting with irregular
pulsation in mostly overtone modes on the low-luminosity AGB to regular
pulsation in fundamental mode of Mira variables 
(cf., e.g., Wood \cite{wood1999}). The pulsation in the 
stellar interior leads to atmospheric motion. Shock fronts reach
the atmosphere, levitate the gas, and lead to a very
extended atmospheric structure and to conditions that are favorable
for the formation of molecules. This leads to a scenario where
molecular layers lie above the continuum-forming photosphere. 
The resulting intensity profiles are very complex and wavelength-dependent
showing features such as multiple components, step-like functions corresponding
to the locations of the shock fronts, or tail-like extensions. The
corresponding visibility profiles deviate from UD models already in the
first lobe of the visibility function, unlike for a regular cool 
giant as described above in Sect.~\ref{sec:RGB}.

Depending on whether or not carbon has been dredged up
from the core into the atmosphere, AGB stars appear in observations to have 
an oxygen-rich or a carbon-rich chemistry. 

\paragraph{Atmospheres of oxygen-rich AGB stars} 
\begin{figure}
\centering
\includegraphics[width=0.6\hsize]{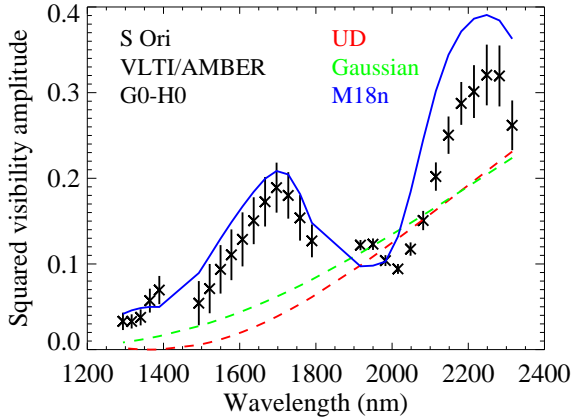}
\caption{VLTI/AMBER observation of the Mira-variable AGB star S~Ori.
The red line denotes the best-fit UD model, the green line the best-fit
Gaussian model, and the blue line the best-fit dynamic model atmosphere
{\it M18n} from Ireland \etal\ (\cite{ireland2004a,ireland2004b}).
From Wittkowski \etal\ (\cite{wittkowski2008}).}
\label{fig:sori}
\end{figure}
Most important molecules
in atmospheres of oxygen-rich AGB stars are TiO, H$_2$O, CO, SiO. 
Self-excited pulsation 
models of oxygen-rich Mira-variable AGB stars (Ireland \etal\ 
\cite{ireland2004a, ireland2004b,ireland2008,ireland2011})
have been successful to describe interferometric 
observations of these sources including their extended atmospheric 
molecular layers (e.g., Woodruff \etal\ \cite{woodruff2004}, Fedele \etal\
\cite{fedele2005},
Woodruff \etal\ \cite{woodruff2009}, Wittkowski \etal\ \cite{wittkowski2011}, 
Hillen \etal\ \cite{hillen2012}).  Fig.~\ref{fig:sori} shows an example 
of an interferometric measurement
of an oxygen-rich Mira-variable AGB star, S~Ori obtained with the 
VLTI/AMBER instrument by Wittkowski \etal\ (\cite{wittkowski2008}).
The bumpy visibility curve is a signature of molecular layers lying above
the photosphere. At wavelength where the molecular opacity is low, we
see the photosphere, the star appears smaller and the visibility larger.
At wavelength where the molecular opacity is larger, wee see the molecular
shell (here most importantly CO and H$_2$O), the star appears larger and
the visibility lower. It is clearly visible that the measured visibility
curve is very different to a UD model already in the first lobe of the
visibility function, unlike in the case of a normal RGB star as
shown in Fig.~\ref{fig:menkar}.
The visibility variations with wavelength resemble reasonably well
the predictions by dynamic model atmospheres. The figure also illustrates
that the VLTI/AMBER instrument is well suited to probe molecular layers
around evolved stars because of its spectro-interferometric capabilities, 
i.e. the simultaneous observation of visibilities at many different bandpasses.
Similar molecular features have subsequently been observed with the 
VLTI/AMBER instrument for oxygen-rich AGB stars of different masses and 
luminosities including three OH/IR stars 
(Ruiz-Velasco \etal\ \cite{ruizvelasco2011}), four additional Mira stars
(Wittkowski \etal\ \cite{wittkowski2011}),
and the intermediate-mass AGB stars RS Cap 
(Mart{\'{\i}}-Vidal \etal\ \cite{marti2011}) 
and $\beta$~Peg (Arroyo-Torres \etal\ \cite{arroyo2014}).
Ohnaka \etal\ (\cite{ohnaka2012,ohnaka2013b}) studied the CO first overtone lines
of the giants BK~Vir and Aldebaran using the high-resolution mode of the VLTI
and found as well evidence for additional CO layers lying
above the model-predicted photospheres.

\begin{figure}
\centering
\includegraphics[width=0.62\hsize]{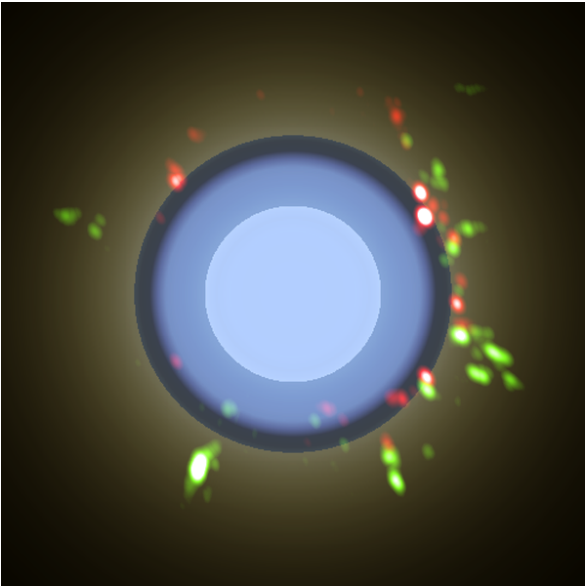}
\caption{VLTI/MIDI and VLBA/SiO maser observations of the 
Mira-variable AGB star S~Ori. The bright blue color indicates
the modeled photosphere, the darker blue color the modeled extended
atmosphere, and the green color the Al$_2$O$_3$ dust shell. 
Overplotted are the reconstructed images of the 42.8\,GHz and 
43.1\,GHz SiO maser emission.
From Wittkowski \etal\ (\cite{wittkowski2007}).}
\label{fig:masers}
\end{figure}
Additional information on the morphology and velocity structure 
of the molecular shells can be obtained by radio observations of
SiO masers that reside within the extended atmospheric layers.
Cotton \etal\ (\cite{cotton2004}), Boboltz \& Wittkowski (\cite{boboltz2005}),
and Wittkowski \etal\ (\cite{wittkowski2007}) combined infrared interferometry
and VLBA observations of SiO masers and found that SiO masers are located
close to two continuum photospheric radii and are co-located with the extended
molecular atmosphere observed by infrared interferometry and possibly with the 
inner Al$_2$O$_3$ dust shell observed by mid-infrared interferometry 
(see Sect.~\ref{sec:dust} below).
The maser spots of S~Ori (Wittkowski \etal\ \cite{wittkowski2007}) showed
a radial gas expansion with a velocity of about 10 km/sec, which is consistent
with velocities predicted by dynamic model atmospheres.
Gray \etal\ (\cite{gray2009}) combined recent dynamic model atmospheres
with a maser propagation code. They found that modeled masers form in rings 
with radii consistent with those found in very long baseline interferometry 
(VLBI) observations and with earlier models. Maser rings, a shock and the 
optically thick layer in the SiO pumping band at 8\,$\mu$m appeared to be 
closely associated.
As an illustration, Fig.~\ref{fig:masers} shows the structure of S~Ori
as derived by Wittkowski \etal\ (\cite{wittkowski2007}) at one epoch,
including the photosphere, the extended molecular layer, 
the Al$_2$O$_3$ dust shell, and the 42.8\,GHz and 43.1\,GHz SiO maser 
emission.

\paragraph{Asymmetries in the atmospheres of oxygen-rich AGB stars}
Ragland \etal\ (\cite{ragland2006,ragland2008}), 
Pluzhnik \etal\ (\cite{pluzhnik2009}), and 
Wittkowski \etal\ (\cite{wittkowski2011}) detected deviations from
point symmetry in the atmospheres of a number of oxygen-rich Mira stars
through the measurement of non-zero closure phases. The spectro-interferometric
capabilities of the VLTI/AMBER instrument used by 
Wittkowski \etal\ (\cite{wittkowski2011})
indicate that the closure phases are strongly wavelength-dependent and 
correlate with the positions of the molecular bands, where larger deviations
from zero closure phases are observed in molecular bands and smaller deviations
at near-continuum bandpasses. This may indicate inhomogeneities or clumps within 
the molecular layers. It is likely that these are relatively small 
inhomogeneities within overall circular shells, as the visibility functions
are fairly well reproduced by centro-symmetric model atmospheres.
Moreover,
the image of the Mira star T Lep reconstructed from VLTI/AMBER data also
shows an overall spherically symmetric shell 
(Le Bouquin \etal\ \cite{lebouquin2009}).

\paragraph{Atmospheres of carbon-rich AGB stars}
\begin{figure}
\centering
\includegraphics[width=0.33\hsize,angle=90]{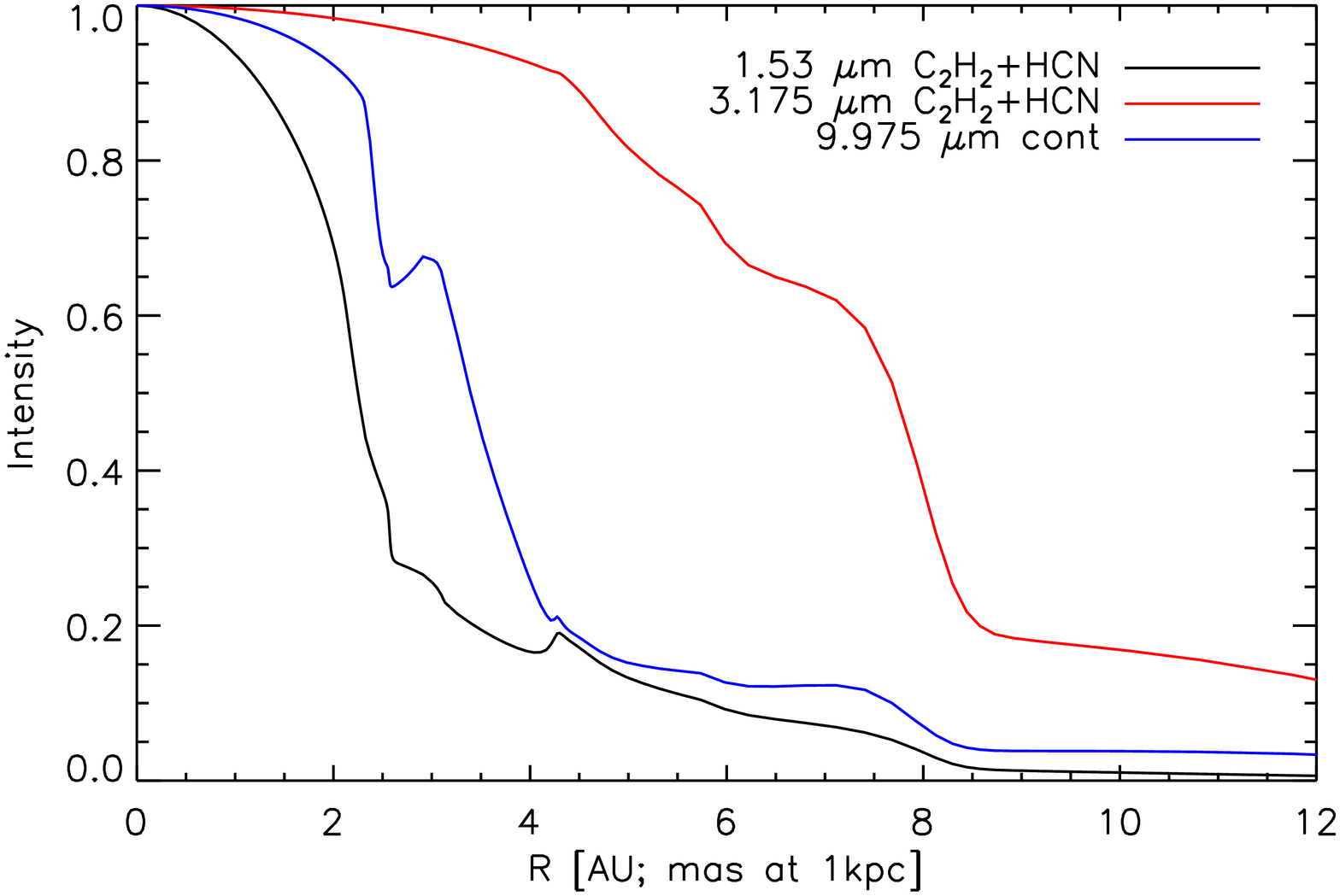}
\includegraphics[width=0.33\hsize,angle=90]{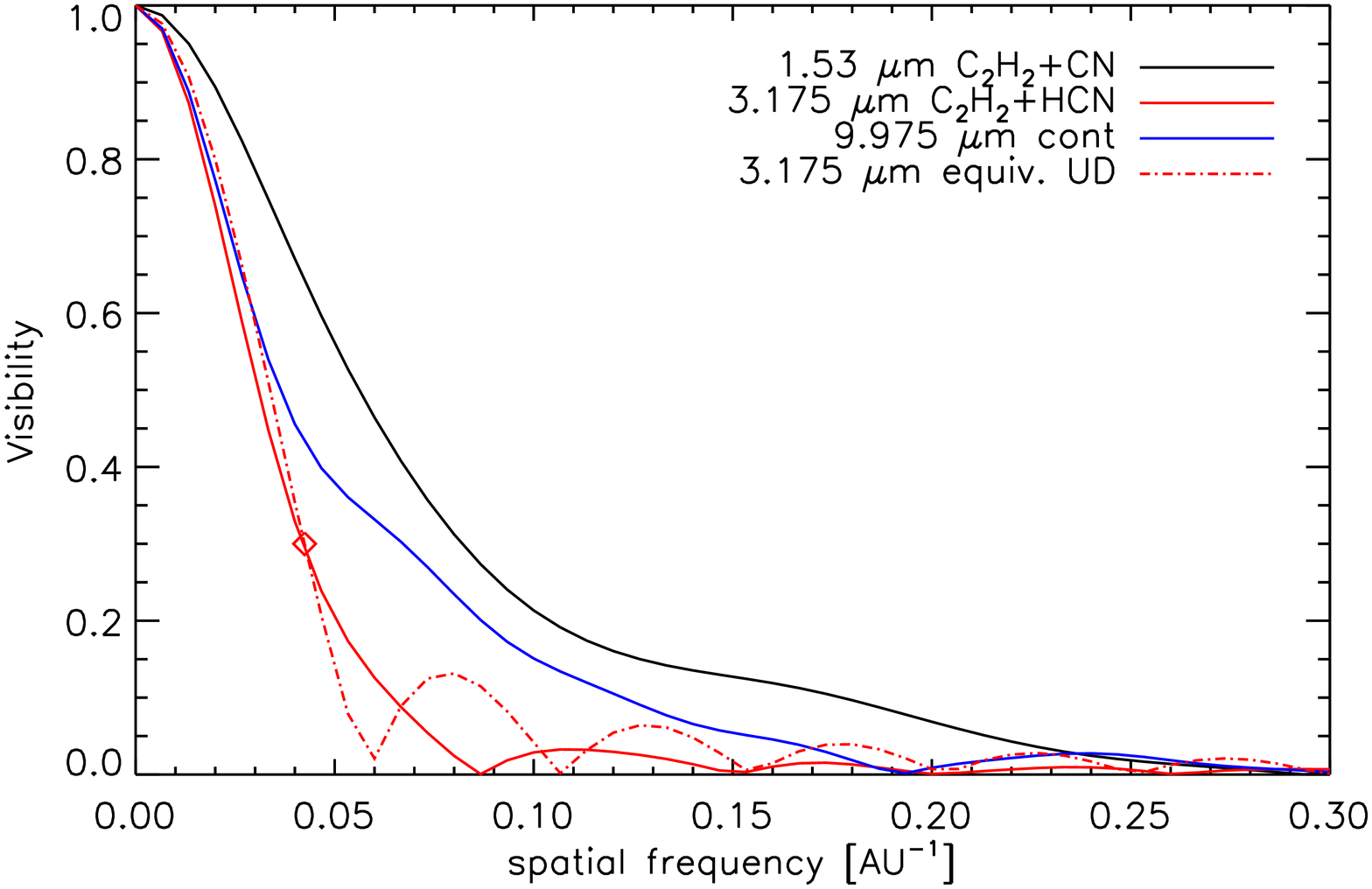}
\caption{Model-predicted intensity profiles and corresponding
visibility curves of carbon-rich AGB stars with mass loss
from Paladini \etal\ (\cite{paladini2009}).}
\label{fig:paladini}
\end{figure}
In carbon-rich stars most of the oxygen is locked in the CO molecule, 
and the atmosphere is enriched of carbon bearing molecules such as 
C$_2$, C$_3$, HCN, C$_2$H$_2$, CH, CN.
Early interferometric studies on C-stars were mainly devoted to stellar 
parameter determination (diameter and effective temperature) and made 
use of geometric models, as dedicated model atmospheres were not available. 
A few examples of these works are Quirrenbach \etal\ (\cite{quirrenbach1994}),
Dyck \etal\ (\cite{dyck1996}), and van Belle \etal\ (\cite{vanbelle1997}).
Model predictions for interferometric observables were presented by 
Paladini \etal\ (\cite{paladini2009}).
They showed that in the near-infrared there is no window to directly measure 
the continuum of carbon stars. This is particularly true for evolved objects
with mass-loss rates above $10^{-8}~M_{\odot}$~yr$^{-1}$.
Models of mass-losing stars exhibit a strong dependency of the 
radius vs. wavelength, and show visibility profiles that are clearly 
different from UDs as one can see in Fig.~\ref{fig:paladini}.
A few examples of comparisons of interferometric data with model atmospheres
of carbon stars can be found in Sacuto \etal\ (\cite{sacuto2011}), 
Paladini \etal\ (\cite{paladini2011}), van Belle \etal\ (\cite{vanbelle2013}),
and Cruzal\'ebes \etal\ (\cite{cruzalebes2013}).
 
\paragraph{Asymmetries in the atmospheres of carbon-rich AGB stars}
As in the case of O-rich AGB stars , the inner envelopes of
carbon stars are also characterized by asymmetric structures
(Ragland \etal\ \cite{ragland2006}, van Belle \etal\ \cite{vanbelle2013}), 
Cruzal\'ebes \etal\ \cite{cruzalebes2013}).
There are various interpretations, including asymmetries driven by 
large convective cells, ellipticity due to stellar rotation,  or
clumps in the molecular layers. 
A possible insight will be gained 
with the new generation of interferometric ''imagers'' at the VLTI
(i.e. the PIONIER, GRAVITY, and MATISSE instruments).

\subsection{Atmospheres of red supergiants}
\label{sec:atmRSG}
It has been observed that photospheres of red supergiants are
surrounded by molecular layers in a similar way as oxygen-rich Miras
(e.g., Danielson \etal\ \cite{danielson1965}, Perrin \etal\ \cite{perrin2005},
Ohnaka \etal\ \cite{ohnaka2009}, Montarg{\`e}s \etal\ \cite{montarges2014}), 
although their pulsation amplitudes
are much lower than for Miras. In a series of papers, 
Wittkowski \etal\ (\cite{wittkowski2012}) and 
Arroyo-Torres \etal\ (\cite{arroyo2013,arroyo2015}) used the VLTI/AMBER
instrument to observe six red supergiants and confirmed
the presence of extended molecular layers for their objects. 
They compared the data to hydrostatic {\tt PHOENIX} model atmospheres
showing that the {\tt PHOENIX} models fit well the spectra, i.e. that
the molecular opacities are included in the model, but that they do not
predict the observed extensions of the molecular layers in the H$_2$O
and CO bands, i.e. that the models are too compact compared to the
observations. Likewise, pulsation models and convection models with 
parameters of RSGs could so far not explain the observed extensions.
Ohnaka \etal\ (\cite{ohnaka2009,ohnaka2011,ohnaka2013a}) studied 
individual CO lines of the red supergiants Betelgeuse and Antares
using the high-spectral resolution mode of the VLTI and found evidence
for large patches of CO gas moving outward and inward with velocities
of up to $\sim$\,30\,km/sec. They also found densities of the outer
atmospheres that are much larger than those predicted by current 
convection models.
At the time of writing this article, it is thus an unsolved problem
which mechanism levitates the atmospheres of RSGs.

It is expected that convection leads to large observable convection cells
at the photospheric layers of RSGs 
(cf. Chiavassa \etal\ \cite{chiavassa2010} and in this book). Imaging studies
with optical interferometers may soon be able to confirm the presence
of such convection cells and to characterize them.

\section{Dust shells and winds from cool evolved stars}
\label{sec:dust}
The dust formation, including the formation of first seed particles and the 
further dust condensation sequence, depends on the chemistry of the 
AGB atmosphere.
\paragraph{Dust shells of oxygen-rich AGB stars}
The details of the dust formation and acceleration processes from oxygen-rich
AGB stars are a matter of current research.
Dust shell parameters were measured in particular by mid-infrared
interferometry (e.g.; Danchi \etal\ \cite{danchi1994}; 
Monnier \etal\ \cite{monnier1997,monnier2000}; 
Weiner \etal\ \cite{weiner2006}; Lopez \etal\ \cite{lopez1997};
Ohnaka \etal\ \cite{ohnaka2005}; Wittkowski \etal \cite{wittkowski2007};
Sacuto \& Chesneau \cite{sacuto2009},
Karovicova \etal\ \cite{karovicova2011,karovicova2013};
Zhao-Geisler \etal\ \cite{zhaogeisler2011,zhaogeisler2012};
Sacuto \etal\ \cite{sacuto2013}).
There are mainly three scenarios that are being discussed, as recently
outlined in more detail by Karovicova \etal\ (\cite{karovicova2013}):
\begin{enumerate}
\item Dust formation starts with TiO clusters,
which can serve as growth centers for both Al$_2$O$_3$ and 
silicates; Al$_2$O$_3$ can also condense on its own with
condensation temperatures around 1400 K (Gail \& Sedlmayer \cite{gail1999}). 
Al$_2$O$_3$ grains may become
coated with silicates at larger radii and can serve as seed 
nuclei for the subsequent silicate formation (e.g. Deguchi \cite{deguchi1980}).
\item Heteromolecular condensation of iron-free magnesium-rich 
(forsterite) silicates based on Mg, SiO, H$_2$O 
(Goumans \& Bromley \cite{goumans2012}). Such
grains may exist at small radii of 1.5 R$_\odot$ to 2 R$_\odot$ 
(Ireland \etal\ \cite{ireland2005}, Norris \etal\ \cite{norris2012}). 
Micron-sized grains may drive the wind (H{\"o}fner \cite{hoefner2008}).
\item SiO cluster formation as seeds for silicate dust formation. 
SiO cluster formation was thought to take place at temperatures 
below 600\,K. However, new
measurements indicate higher SiO condensation temperatures, and may
be compatible with observed dust temperatures around 1000\,K 
(Gail \etal\ \cite{gail2013}).
\end{enumerate}
Karovicova \etal\ (\cite{karovicova2013}) favor scenario (1), as
the combination of Al$_2$O$_3$ and warm silicate grains fits well 
observed spectra and visibilities. However, additional dust grains
with relatively flat spectra are needed to explain the observed 
optical depths for some of their targets. Sacuto \etal\ (\cite{sacuto2013})
favor scenario (2), but require the addition of a significant  amount
of Al$_2$O$_3$ grains to fit the observed visibility spectra.
Taking both results together, Al$_2$O$_3$ grains and iron-free 
magnesium-rich (forsterite) silicates may co-exist at small radii.

\paragraph{Dust shells of carbon-rich AGB stars}
Two different dust species are observed in the atmospheres of carbon-rich 
AGB stars: amorphous carbon, and silicon carbide (SiC). The amorphous carbon 
is feature-less in the spectrum, and also in the visibilities.
Interferometrically speaking, the presence of amorphous carbon can be 
recognized by comparing the diameter of the star at different wavelengths.
If an amorphous dust opacity is present, the star will appear larger 
(i.e. lower visibilities) at long wavelengths with respect to the
near infrared diameter. The SiC dust appears with a more or less pronounced 
emission in the spectrum centered at 11.3~$\mu$m.
In the visibility profile the presence of SiC will increase the size
of the star at the wavelength of this feature with respect to 
wavelengths between 9\,$\mu$m and 10\,$\mu$m. 
The wind in the atmosphere of carbon AGB stars is driven by 
radiative pressure on small dust particles. This scenario was tested 
interferometrically by comparing self-consistent 
model atmospheres including a dust driven stellar wind with interferometric 
observations (Sacuto \etal\ \cite{sacuto2011}).
Recent interferometric studies of dust shells in carbon star were presented 
by Ohnaka \etal\ (\cite{ohnaka2007}), 
Zhao-Geisler \etal\ (\cite{zhaogeisler2012}),  and 
Ladjal (\cite{ladjal2011}).
These authors report interferometrical variability in the $N-$band, 
and identify a wide region for the dust formation zone that ranges from 
5 up to 18 stellar radii.

\paragraph{Dust shells of red supergiants}
Red supergiants show mass-loss rates comparable to and exceeding those
of oxygen-rich AGB stars (e.g. de Beck \etal\ \cite{debeck2010}).
There are also indications that their dust formation process and dust condensation
sequence may in principle be similar as well 
(Verhoelst \etal\ \cite{verhoelst2006,verhoelst2009},
Perrin \etal\ \cite{perrin2007}), although the corresponding process is 
not yet understood (cf. also Sect.~\ref{sec:atmRSG} above).
Ohnaka \etal\ (\cite{ohnaka2008b}) studied the dust envelope of the 
red supergiant WOH G64 in the Large Magellanic Cloud with the VLTI/MIDI
instrument and detected an optically and geometrically thick silicate torus,
a result that also lead to a refinement of the stellar parameters of this
source. The mass-loss process, in particular the mass-loss rates
and the geometry of the circumstellar dust shell, is important for
the different paths of further evolution of massive evolved stars
toward the different types of core-collapse SNe (e.g. Groh et al. 
\cite{groh2013}).
\section{Asymmetries in the dusty region of AGB stars}
One of the hot topics in the field of evolved stars over the last few decades 
concerns the geometry of the mass-loss process, and how it influences the 
following phase of stellar evolution when the star expels its envelope as a 
PN.
Although $\approx 70\%$ of the observed PNe are not spherically symmetric
(e.g. Miszalski \etal\ \cite{miszalski2009}), and neither are the winds of 
post-AGB objects (e.g. Bujarrabal \etal\ \cite{bujarrabal2013}),
the AGB wind morphology is widely regarded as such. So far, the most accepted 
explanation that accounts for the asymmetries observed in the PN phase is the 
presence of a nearby companion in the environment of the (once) mass-losing 
star. Observations of an interaction between a companion and stellar wind 
on the AGB were only successful for a handful of objects 
(Mauron \& Huggins \cite{mauron2006}, Dinh-V.-Trung \& Lim \cite{dinh2008}, 
Mayer \etal\ \cite{mayer2011}, 
Maercker \etal\ \cite{maercker2012}, 
Mayer \etal\ \cite{mayer2013}). This implies that either there are plenty of 
undiscovered binary stars hidden in the envelope of AGBs, or the asymmetries 
on the following evolutionary stage are caused by the stellar wind.

Infrared interferometry is a prime technique to provide the closest glimpse 
to the onset of the stellar wind,
where big scale asymmetries induced by the mass-loss process could form.
Image reconstruction may be the easiest way to  detect asymmetries, 
but unfortunately this is not yet routine when dealing with interferometry.
Asymmetries have so far been detected by direct investigations
of visibility data in the Fourier plane.
There are two ways to detect departures from symmetry with interferometry 
in the Fourier plane: (i) by comparing visibilities observed
with the same baseline lengths (and at the same epoch for variable targets
such as AGB stars), but different position angles;
(ii) by detecting a non zero phase signature.
When we speak about phase we need to distinguish between closure and 
differential phase.
The closure phase was extensively explained in a previous
VLTI school by Monnier (\cite{monnier2007}).
The differential phase is defined as the phase difference between 
different spectral channels. In this contribution we will mention the 
differential phase that is provided by MIDI, which has been
explained by, e.g., Ohnaka \etal\ (\cite{ohnaka2008a}) or recently by
Tristram \etal\ (\cite{tristram2014}).

\begin{figure}
\centering
\includegraphics[width=0.62\hsize,angle=-90]{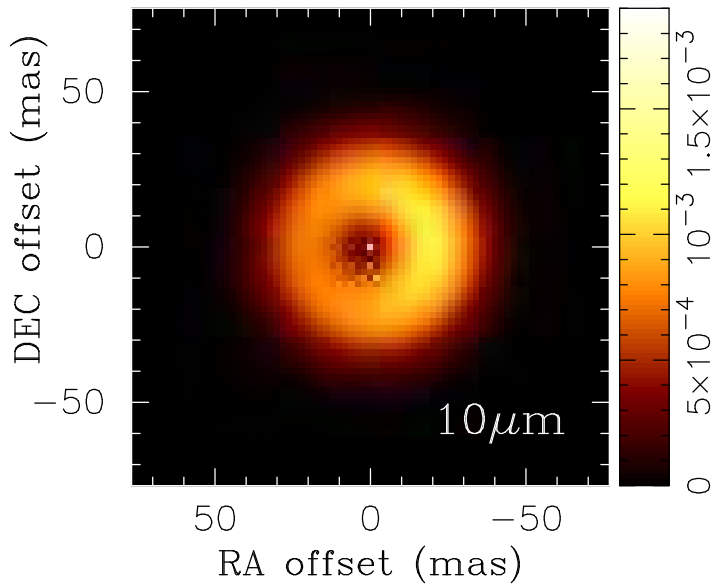}
\caption{Model of the best-fit asymmetric-disk model of the silicate carbon
star BM Gem at 10\,$\mu$m from Ohnaka \etal\ (\cite{ohnaka2008a}).}
\label{fig:ohnaka}
\end{figure}

Over the last decade few asymmetries were reported on the AGB, but the 
interpretation is very difficult, as one would expect.
Tatebe \etal\ (\cite{tatebe2006}) reported the detection of asymmetries 
in the dust region
of six M- and S-type AGB stars. The asymmetry seen in R~Aqr was ascribed to 
binary interaction, while the origin of the asymmetries in the other stars 
remained unclear.
It is known from direct imaging that the innermost region of the prototype 
of carbon stars IRC+10216 is clumpy. This was confirmed with non zero closure 
phase observations with ISI by Chandler \etal\ (\cite{chandler2007}).
Even in this well studied case, it is not clear if the clumps follow a 
random direction or a preferential one (like disc or spiral). This is mainly 
due to the small $uv$-coverage of the observations.
Two J-type stars were observed by VLTI/MIDI, and showed a clear signature 
of departure from symmetry in the differential phase
(Deroo \etal\ \cite{deroo2007}; Ohnaka \etal\ \cite{ohnaka2008a}).
A general consensus about these objects is that they are binaries with 
an unseen companion, probably a main sequence star (Morris \cite{morris1987};
Lloyd Evans \cite{lloydevans1990}, and Yamamura \etal\ \cite{yamamura2000}).
The observations were interpreted as circum-binary or circum-companion discs,
in which the oxygen-material has been stored while the AGB star transformed
from an oxygen-rich star to a carbon-rich star (cf. also Ohnaka \etal\
\cite{ohnaka2006}).
As an illustration, Fig.~\ref{fig:ohnaka} shows the best-fit asymmetric 
ring model of the silicate carbon star BM Gem at a wavelength of 10\,$\mu$m
from Ohnaka \etal\ (\cite{ohnaka2008a}). This ring model can be 
characterized by a
broad ring with a bright region offset from the unresolved star, and can
be interpreted as a system with a circum-companion disk.
We note that the shape of the differential phase
for the two stars by Deroo \etal\ and Ohnaka \etal\ is very different. 
This might be the effect of the different chemistry, mixed with the 
effect of different geometries.

A similar signature was detected for the two Mira variables  
R~For and RT~Vir (Paladini \etal\ \cite{paladini2012}; 
Sacuto \etal\ \cite{sacuto2013}).
The asymmetry could be interpreted as a clump, without involving a 
disc structure.
The authors were able to reproduce the visibility and differential phase 
with a simple model composed by a UD representing the central star, 
a Gaussian for the extended envelope, and a Dirac function representing 
a bright dust clump.
The shape of the differential phase of R~For was matching the one observed 
by Deroo \etal\ (\cite{deroo2007}).
An asymmetry was also found in the dusty region of the O-rich star with a 
peculiar CO profile SV Psc by Klotz \etal\ (\cite{klotz2012}).
In this case the asymmetry is again interpreted as the presence of a 
disc or of a companion.
Again the poor $uv$-coverage makes data interpretation complex and at this 
stage it is not possible to conclude if disks are present at this 
spatial scale, and if they are associated with hidden unknown binary 
companions, or with the mass loss mechanism.
Interestingly, the number of asymmetries detected in AGB stars with 
optical interferometry is rather small compared to the number of 
papers published.
Ohnaka \etal\ (\cite{ohnaka2005}), 
Wittkowski \etal\ (\cite{wittkowski2007}), 
Sacuto \etal\ (\cite{sacuto2008,sacuto2011}), Zhao-Geisler \etal\ 
(\cite{zhaogeisler2011},\cite{zhaogeisler2012}),
Karovicova \etal\ (\cite{karovicova2011,karovicova2013}), 
Klotz \etal\ (\cite{klotz2013}) did not detect 
any asymmetries of the circumstellar dust shells.
This might imply that the asymmetric structures detected with MIDI are 
not very common among the AGBs. Alternatively, they might appear
only at low visibility values (below 0.1), while most of the observations 
presented so far did not sample this part of the visibility curve.


\end{document}